# The University of California San Francisco, Brain Metastases Stereotactic Radiosurgery (UCSF-BMSR) MRI Dataset

Jeffrey D. Rudie MD PhD[1,2], Rachit Saluja MSE[3], David A. Weiss MSE[1], Pierre Nedelec MS[1], Evan Calabrese MD PhD[1,4], John B. Colby MD PhD[1], Benjamin Laguna MD[1], John Mongan MD PhD[1], Steve Braunstein MD PhD[4], Christopher P. Hess MD PhD[1], Andreas M. Rauschecker MD PhD[1], Leo P. Sugrue MD PhD[1], Javier E. Villanueva-Meyer MD[1]

1. University of California San Francisco, Department of Radiology & Biomedical Imaging
2. University of California, San Diego, Department of Radiology
3. University of Pennsylvania, Department of Radiology
4. Duke University, Department of Radiology
5. University of California San Francisco, Department of Radiation Oncology

**Corresponding Author**
Jeff Rudie, MD, PhD
Jeff.Rudie@gmail.com

**Abstract**

The University of California San Francisco Brain Metastases Stereotactic Radiosurgery (UCSF-BMSR) dataset is a public, clinical, multimodal brain MRI dataset consisting of 560 brain MRIs from 412 patients with expert annotations of 5136 brain metastases. Data consists of registered and skull stripped T1 post-contrast, T1 pre-contrast, FLAIR and subtraction (T1 pre-contrast - T1 post-contrast) images and voxelwise segmentations of enhancing brain metastases in NifTI format. The dataset also includes patient demographics, surgical status and primary cancer types. The UCSF-BSMR has been made publicly available in the hopes that researchers will use these data to push the boundaries of AI applications for brain metastases.

## Introduction

Public datasets, such as those made available through The Cancer Imaging Archive (TCIA; (1)) and multimodal Brain Tumor Segmentation challenges (BraTS; (2)) have been critical in supporting advances in the field of biomedical image segmentation in neuro-oncology, particularly in glioma. Brain metastases are the most common central nervous system tumor (3,4), and new effective treatment methods have led to a dramatic rise in imaging for their detection (5). Detection of very small metastases is a major challenge (6–8) that can potentially be overcome by public data sharing initiatives (9).

Here we present the University of California, San Francisco Brain Metastases Stereotactic Radiosurgery (UCSF-BMSR) MRI Dataset. This dataset contains 560 multimodal brain MRIs with expert annotations of brain metastases in 412 patients undergoing gamma knife radiosurgery. There are voxelwise annotations of 5136 brain metastases, including two sets of annotations in 99 patients. We also provide a current benchmark of metastasis detection and segmentation performance in this dataset using the nnU-Net (10).

## Materials and Methods

*Patient Sample*

Data collection was performed in accordance with relevant guidelines and regulations and was approved by the UCSF institutional review board with a waiver for consent. The dataset consists of 560 brain MRIs from 412 patients (mean age, 61 ± [SD] 12 years, 238 female, 174 male; Table 1) who were undergoing stereotactic radiosurgery planning at the UCSF medical center. MRIs were identified through a retrospective search of institutional radiology archives (mPower, Nuance Communications, Inc) of stereotactic radiosurgery studies performed between January 1st 2017 and February 29th, 2020. Exclusion criteria from initially selected scans were patients without enhancing intracranial metastases ($n$ = 26), only dural-based or leptomeningeal metastases ($n$ = 9), or missing sequences/corrupted data ($n$ = 8). This dataset overlaps with a prior study (8), but also includes the fluid-attenuated inversion recovery (FLAIR) sequence and skull stripping, noting that three examinations from that dataset were not included here given missing FLAIR. Prior craniotomies for resections or biopsies were present in 137 (24.5%) of the scans, noting that none of the patients used in the test set (n = 99) had craniotomies. Of the total pre-operative training images, 324 are included as an external dataset in the 2023 American Society of Neuroradiology – Medical Image Computing and Computer Assisted Intervention Society (ASNR-MICCAI) BraTS challenge (https://www.synapse.org/#!Synapse:syn51156910/wiki/622553; (9) with corresponding subject IDs in **Table S1**).

*Imaging Data Acquisition*

T1-weighted spoiled gradient-echo (T1), T1 post-contrast (T1-post) images, and 2D post-contrast T2 FLAIR images are included for each scan. The majority (373; 66%) of studies were acquired on a 1.5 Tesla GE SignaHDxt scanner, 149 (27%) were acquired on a 1.5 Tesla Philips Achieva scanner, and 38 (7%) were acquired on a 3.0 Tesla GE Discovery MR750. The scanner type, Tesla strength, matrix size, in-plane axial dimension and slice thickness for each scan are available in **Table S1**.

Although acquisition parameters varied slightly, representative values from a 1.5 Tesla GE SignaHDxt scanner for the T1 and T1-post images were as follows: repetition time, 8.8 msec; echo time, 3.3 msec; inversion time, 450 msec, flip angle, 11°; matrix size, 256 x 256 x 106; and voxel size, 0.71 x 0.71 x 1.5 mm. Across all acquisitions, the T1-post image in-plane axial voxel dimension was less than 1 x 1 mm with a slice thickness of less than or equal to 1.5 mm in 498 images (89%). Representative values for the 2D post-contrast FLAIR sequence were as follows: repetition time, 9175 msec; echo time, 109 msec; inversion time, 2600 msec; matrix size, 256 x 256 x 38; and voxel size, 0.98 x 0.98 x 5 mm.

*Brain Metastases Annotations*

Reference standard brain metastasis voxelwise segmentations were created using ITK-SNAP (11); http://www.itksnap.org/) by one of two neuroradiology fellows (J.D.R. and B.L.) or one of two attending neuroradiologists (J.E.V. and L.P.S; with 5 and 4 years of experience as neuroradiology attendings). The associated final radiology report, containing a description of all the metastases, were used as a reference during the annotation process. The T1, T1-post, and subtraction images were available to guide the manual segmentations. The enhancing portions of all metastases were segmented. Areas of T1 intrinsic hyperintensity and central necrosis were not included in the segmentation masks. In studies with resection cavities, the nodular enhancing portion of the resection cavities were included in the segmentation masks. Ninety nine of the images, representing unique individuals previously used as the test set in (8), were segmented by both attending neuroradiologists (J.E.V. and L.P.S) without reference to the final radiology report. Final reference standard segmentations for these test images were generated by combining and refining the two segmentations with reference to the final radiology report.

*Image Pre-Processing*

Images were first converted from Digital Imaging and Communications in Medicine, or DICOM, format into NifTI format using dcm2niix. The T1 and FLAIR images were registered into the T1-post space by rigid registration (6 degrees of freedom) using the FMRIB Software Library tool, FLIRT (12). Subtraction images were generated by subtracting the registered T1 from T1-post images using fslmaths (12). For the purposes of protecting personal health information, all images were skull stripped using an in-house T1-post skull-stripping nnU-Net skull stripping network (https://github.com/rachitsaluja/UCSF-BMSR-benchmarks). No other image normalization, scaling/resizing or atlas registration was performed prior to input into the neural network.

*nnU-Net Convolutional Neural Network Experiments*

In order to evaluate current state-of-the-art performance of metastasis segmentation and detection, we performed several experiments using the nnU-Net (10), a 3D fully-convolutional network (3D Res-U-Net). The nnU-Net is a self-configuring method that performs pre-processing (including z-score normalization, resampling, augmentation) and network architecture and hyperparameter tuning. We evaluated different multichannel input combinations of the T1, T1-post, subtraction, and FLAIR images, a larger batch size (batch size of 6 rather than default of 2), pretraining the network on the BraTS 2021 training dataset for enhancing tumor (n=1251), and the nnU-net modifications used by the winners of the BraTS 2021 challenge (2,13), which included group normalization and a larger network size. We also evaluated performance with different training sample sizes. Performance was evaluated in the 99 test images for each of these experiments. In all experiments, the 3D patch size was auto selected to be 64x192x160 (x, y, z dimensions), and training was performed on a RTX 3090 GPU (CUDA version 11.2; NVIDIA corporation, Santa Clara, CA; 24 GB memory) for 1000 epochs using a combination of cross entropy and Dice loss function (1:1). We also calculated performance metrics between the two expert segmentations in the 99 test images relative to each other and relative to the final reference standard segmentations. Code used to for the nnU-Net implementation and pretrained networks are available at (https://github.com/rachitsaluja/UCSF-BMSR-benchmarks). Statistical significance between algorithm and radiologist performance was not assessed.

## Results

### *Size and Distribution of Brain Metastases*

**Table 1** provides an overview of patient characteristics and metastasis number and size distribution, with details for each case provided in **Table S1**. The most common types of primary cancers were lung, breast, and melanoma. A total of 5136 brain metastases were segmented in the 560 MRIs, with a mean of 9.2 +/- 12.9 metastases and a median of 5 (IQR 2-11) metastases. The mean and median total metastasis volumes per MRI were 5.3 +/- 6.9 $cm^3$ and 2.6 $cm^3$ (IQR, 0.6-6.9 $cm^3$). The mean and median individual metastasis volumes were 0.57 +/- 2.1 $cm^3$ and 0.05 $cm^3$ (IQR, 0.02-0.18 $cm^3$). The distribution of metastases per scan and metastasis volumes are shown in **Figure 1**.

### *Performance of nnU-Net and Radiologist Segmentations*

**Table 2** provides the segmentation performance on the test set for the experiments detailed in the Methods section. Performance metrics included mean and median Dice, overall sample and average sample sensitivity, precision and F1 scores (Lesion wise Dice). While there were minimal differences between different multichannel inputs, the model with T1, T1-post and subtraction images had the highest overall sensitivity (82.91%), so it was used for the remaining experiments. The additional modifications did not significantly alter performance, noting that the winning BraTS 2021 modifications (13) did have the highest overall F1 score, and the model pretrained on BraTS 2021 enhancing tumor dataset had slightly lower performance. Dice scores generally plateaued at ~200 training samples, but sensitivity and F1 scores did continue to improve minimally with larger sample sizes. Finally, we found that the nnU-Net yielded slightly Dice scores (median Dice, 0.89) that were higher relative to inter-rater reliability of segmentations between of the two radiologists (median Dice, 0.83) and prior work (median Dice, 0.75 (8)), but yielded sensitivity scores that were lower than that of radiologists (83% vs 86-89%).

*Data and Code Availability*

Currently, 461 MRIs and annotations from the training set are available to download under a non-commercial license at https://imagingdatasets.ucsf.edu/dataset/1. Data from the 99 individuals in the test set will become available after the completion of the 2024 MICCAI (Medical Image and Computing and Computer Assisted Intervention Society) challenge. Code used to for the nnU-Net implementation and pre-trained networks are available at (https://github.com/rachitsaluja/UCSF-BMSR-benchmarks).

**Discussion**

Automated detection and segmentation of brain metastases represents an ideal use case for the translation of artificial intelligence methods into clinical practice. Here, we present the publicly available UCSF-BMSR MRI Dataset, which contains 560 multimodal MRIs with expert-annotations of over 5000 metastases and associated clinical and image information.

Other publicly available brain metastases datasets include Stanford BrainMetShare (https://aimi.stanford.edu/brainmetshare; (6)), NYUmets (https://nyumets.org/; (14)), and the mathematical oncology lab (MOLAB) brain metastases dataset ((15) https://molab.es/datasets-brain-metastasis-1). The UCSF-BMSR contains the largest number of annotated brain metastases. The Stanford BrainMetShare contains 105 scans with annotations of 1517 metastases. While the NYUmets is the largest dataset with >8000 studies, only a subset of 2367 metastases are segmented as part of radiation treatment planning. The MOLAB dataset contains 637 MRIs but only has 593 "semi-automatic" segmentations available, which may be missing small metastases.

Data sharing efforts and competitions, such as those by BraTS and the Radiological Society of North America, have been critical for the advancement of biomedical image segmentation tasks as they provide large amounts of annotated data and a performance benchmark. The currently in progress 2023 ASNR-MICCAI BraTS challenge on brain metastases (9) includes data from UCSF-BMSR, Stanford BrainMetShare, NYUmets and several other sites. Larger and more heterogeneous data should allow for improved algorithm performance and generalizability, particularly for difficult to detect small metastases, as well as a clearer comparison of different algorithms.

In the UCSF-BMSR MRI dataset, we show that segmentation performance using the nnU-net (median Dice of 0.89) was at the level of neuroradiologist inter-rater reliability (median Dice of 0.83) and better than the previously reported network on this dataset (median Dice, 0.75; (8)). However, sensitivity (83%) was still lower than neuroradiologists (86%-89%). Adding the post-contrast FLAIR sequence did not improve sensitivity despite consensus recommendations (16), noting that this dataset included 2D FLAIR images acquired on 1.5 Tesla scanners as part of the radiosurgery protocol. We did find that sensitivity continued to improve with larger sample sizes, suggesting the possibility that larger public datasets like the 2023 ASNR-MICCAI BraTS challenge and/or novel network architectures/loss functions will allow for continued performance gains. This will be critical in helping lead to clinical implementation of automated brain metastasis segmentation algorithms.

In summary, the integration of artificial intelligence tools into clinical workflows should allow for more rapid and precise quantitative assessments of disease burden, ultimately enabling more accurate and efficient diagnosis and treatment. The evaluation and treatment of brain metastases represents an excellent use case for which more public data are needed. We hope that releasing this dataset will lead to the development of algorithms with better segmentation performance and ultimately improved care of patients with intracranial metastatic disease.

## Tables

**Table 1. Patient Demographics, Cancer Types, and Metastases Statistics**

| | Total |
|---|---|
| **Demographics** | |
| Number of Patients | 412 |
| Number of MRIs | 560 |
| Age (y) | 61 +/- 12 |
| Males:Females | 174:238 (43%:57%) |
| **Primary Cancer Types** | |
| Lung | 172 (41.7%) |
| Breast | 96 (23.3%) |
| Melanoma | 52 (12.6%) |
| Renal | 22 (5.3%) |
| Head and Neck | 11 (2.6%) |
| Other Genitourinary | 10 (2.4%) |
| Other Gastrointestinal | 9 (2.2%) |
| Rectal | 8 (1.9%) |
| Neuroendocrine | 5 (1.2%) |
| Colon | 6 (1.5%) |
| Prostate | 10 (2.4%) |
| Thyroid | 4 (1.0%) |
| Other/Unknown | 7 (1.7%) |
| **Metastases Information** | |
| Prior Craniotomy/Resection/Biopsy:No Prior Surgery | 137:423 (24.4%:75.5%) |
| Total Number of Metastases | 5136 |
| Number of Metastases | 9.2 ± 12.9, 5 (2-10) |
| Total Metastases Volume ($cm^3$) | 5.3 ± 6.9, 2.6 (0.6-6.9) |
| Individual Metastasis Volume ($cm^3$) | 0.57 ± 2.1, 0.05 (0.02-0.18) |

Note.— Continuous variables are shown as mean ± standard deviation (SD), with median (IQR) also reported for metastases variables. Categorical variables are reported as number (percentage).

## Table 2. nnU-Net Segmentation Performance Metrics

| Experiment | Dice | | Overall Detection Performance | | | Average Detection Performance | | |
|---|---|---|---|---|---|---|---|---|
| **Different Inputs** | **Mean +/- SD** | **Median (25-75 IQR)** | **Sensitivity** | **Precision** | **F1 score** | **Sensitivity** | **Precision** | **F1 score** |
| T1-post | 0.84 ± 0.12 | 0.88 (0.82-0.91) | 81.92% | 91.77% | 0.87 | 88.2 ± 17.7% | 93.6 ± 14.3% | 0.87 ± 0.19 |
| Subtraction | 0.85 ± 0.14 | 0.89 (0.83-0.92) | 78.67% | 89.41% | 0.84 | 86.9 ± 20.9% | 89.8 ± 19.3% | 0.85 ± 0.21 |
| T1 and T1-post | 0.86 ± 0.11 | 0.89 (0.84-0.92) | 81.21% | 92.74% | 0.87 | 88.3 ± 17.2% | 94.1 ± 13.3% | 0.88 ± 0.18 |
| *T1, T1-post, and Subtraction* | *0.86 ± 0.11* | *0.89 (0.85-0.92)* | *82.91%* | *93.17%* | *0.88* | *89.3 ± 17.0%* | *94.8 ± 13.4%* | *0.88 ± 0.19* |
| T1, T1-post, and FLAIR | 0.86 ± 0.10 | 0.89 (0.84-0.91) | 80.79% | 93.77% | 0.87 | 88.4 ± 18.0% | 95.1 ± 12.7% | 0.88 ± 0.19 |
| T1, T1-post, Subtraction, and FLAIR | 0.86 ± 0.10 | 0.89 (0.85-0.92) | 81.36% | 93.35% | 0.87 | 88.9 ± 17.2% | 95.1 ± 11.7% | 0.88 ± 0.18 |
| **Different Network Configurations** | | | | | | | | |
| Batch Size = 6 | 0.85 ± 0.11 | 0.89 (0.84-0.92) | 81.21% | 95.36% | 0.88 | 87.0 ± 19.2% | 96.4 ± 10.9% | 0.88 ± 0.19 |
| Five-Fold Cross Validation | 0.86 ± 0.11 | 0.89 (0.84-0.92) | 81.07% | 93.49% | 0.87 | 88.1 ± 18.2% | 94.9 ± 13.9% | 0.88 ± 0.19 |
| Pre-trained on BraTS 2021 | 0.84 ± 0.12 | 0.87 (0.81-0.91) | 76.13% | 93.58% | 0.84 | 84.4 ± 21.9% | 94.0 ± 17.1% | 0.86 ± 0.20 |
| Luu et al., 2021 Configuration | 0.85 ± 0.13 | 0.89 (0.84-0.92) | 81.78% | 94.76% | 0.88 | 88.9 ± 17.3% | 95.8 ± 12.2% | 0.89 ± 0.18 |
| **Smaller Training Sample Sizes** | | | | | | | | |
| n=10 | 0.72 ± 0.22 | 0.79 (0.65-0.87) | 60.03% | 97.25% | 0.74 | 73.5 ± 28.0% | 95.6 ± 16.6% | 0.80 ± 0.23 |
| n=25 | 0.76 ± 0.21 | 0.84 (0.71-0.90) | 61.72% | 97.98% | 0.76 | 74.5 ± 28.4% | 96.1 ± 15.8% | 0.81 ± 0.23 |
| n=50 | 0.82 ± 0.15 | 0.86 (0.79-0.90) | 67.37% | 97.15% | 0.80 | 78.8 ± 25.8% | 95.9 ± 14.5% | 0.83 ± 0.22 |
| n=75 | 0.82 ± 0.16 | 0.87 (0.80-0.90) | 73.59% | 95.42% | 0.83 | 82.3 ± 23.1% | 95.4 ± 15.2% | 0.86 ± 0.20 |
| n=100 | 0.82 ± 0.16 | 0.88 (0.79-0.91) | 74.44% | 92.62% | 0.83 | 82.6 ± 22.8% | 93.2 ± 18.4% | 0.85 ± 0.21 |
| n=150 | 0.83 ± 0.15 | 0.87 (0.81-0.91) | 74.15% | 94.59% | 0.83 | 82.7 ± 22.9% | 94.1 ± 17.6% | 0.84 ± 0.20 |
| n=200 | 0.84 ± 0.15 | 0.88 (0.81-0.92) | 78.11% | 95.34% | 0.86 | 86.3 ± 19.4% | 96.3 ± 11.0% | 0.88 ± 0.19 |
| n=300 | 0.85 ± 0.12 | 0.89 (0.84-0.92) | 80.93% | 93.17% | 0.87 | 87.7 ± 19.0% | 94.5 ± 13.3% | 0.87 ± 0.19 |
| n=400 | 0.85 ± 0.12 | 0.89 (0.84-0.92) | 81.64% | 93.08% | 0.87 | 88.7 ± 21.9% | 93.8 ± 13.4% | 0.88 ± 0.19 |
| **Human Inter-Rater Performance** | | | | | | | | |
| Radiologist 1 vs Radiologist 2 | 0.83 ± 0.09 | 0.86 (0.80-0.89) | 86.71% | 85.36% | 0.86 | 90.7 ± 15.7% | 88.5 ± 21.2% | 0.87 ± 0.19 |
| Radiologist 2 vs Radiologist 1 | 0.83 ± 0.09 | 0.86 (0.80-0.89) | 84.68% | 87.87% | 0.86 | 89.8 ± 17.9% | 89.4 ± 20.0% | 0.87 ± 0.20 |
| Radiologist 1 vs Reference Standard | 0.90 ± 0.06 | 0.91 (0.89-0.94) | 85.59% | 97.15% | 0.92 | 90.4 ± 16.8% | 97.9 ± 14.4% | 0.92 ± 0.17 |
| Radiologist 2 vs Reference Standard | 0.94 ± 0.06 | 0.96 (0.93-0.98) | 88.56% | 95.42% | 0.94 | 92.0 ± 14.2% | 97.9 ± 14.4% | 0.93 ± 0.16 |

Note.— The different network configurations and smaller training sample sizes were all done with T1, T1-post and Subtraction inputs. Radiologists 1 and 2 represent different attending neuroradiologists. The reference standard represents consensus of the two neuroradiologists segmentations with reference to the final radiology report. BraTS = Brain Tumor Segmentation dataset, FLAIR = fluid-attenuated inversion recovery

**Figures**

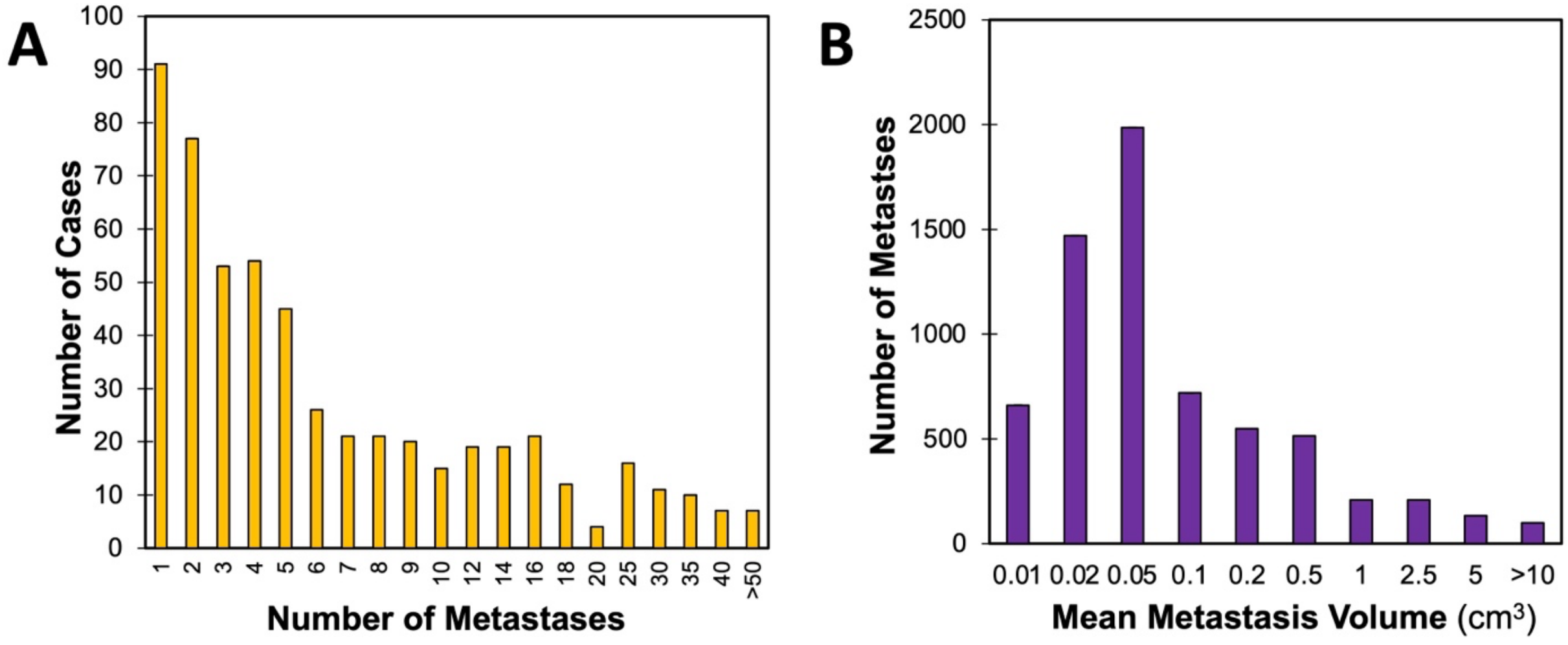

**Figure 1.** Distribution of number and size of brain metastases. Counts of number of manually segmented brain metastases per MRI (a) and average brain metastasis volume (b).

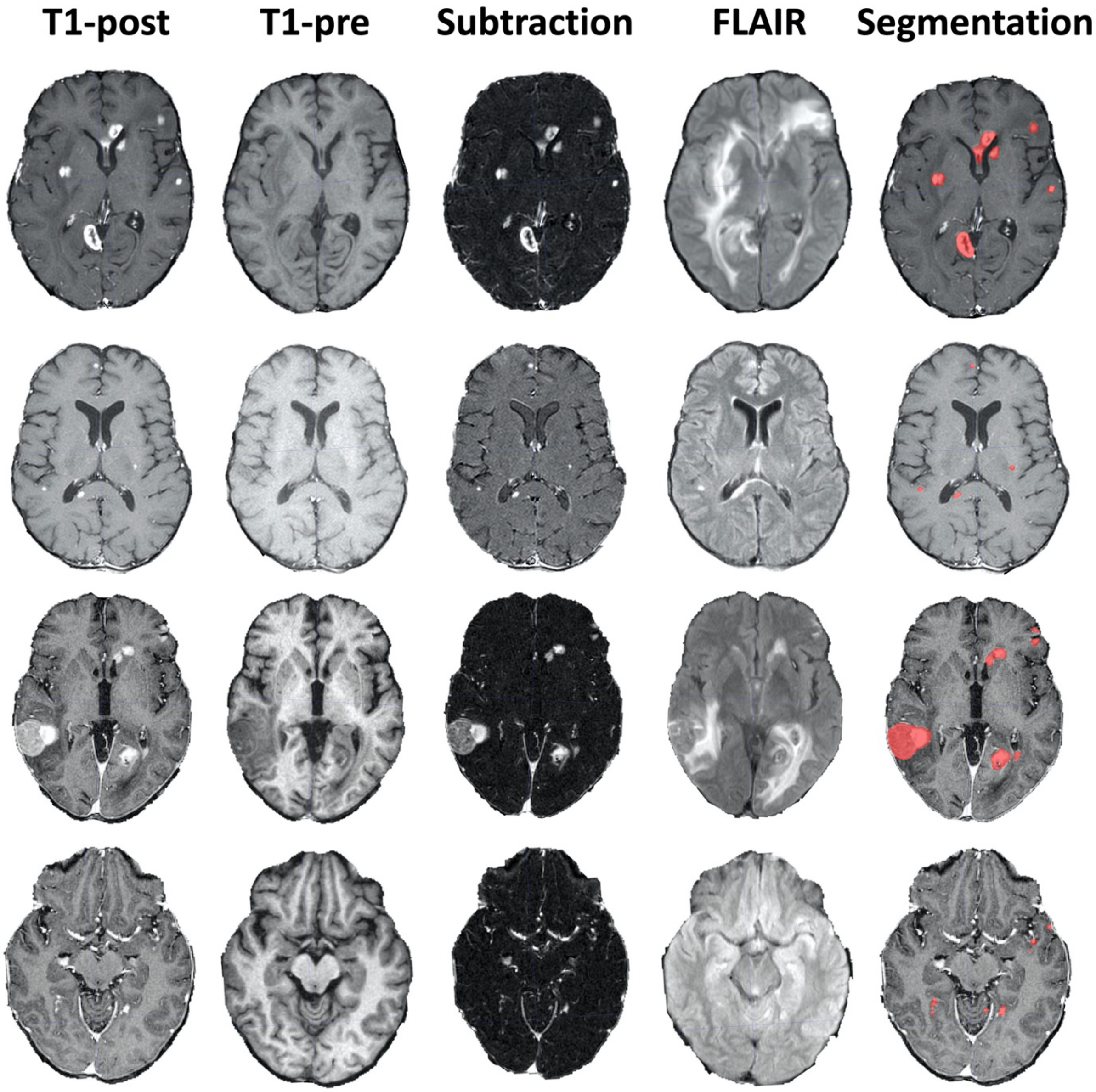

**Figure 2.** Example Brain Metastases Data. Four example MRIs with axial T1 post-contrast images (T1-post), T1 pre-contrast images (T1-pre), subtraction images, FLAIR images, and ground truth segmentations overlaid on the T1 post-contrast images.